\pdfoutput=1
\documentclass[5p,pdftex]{elsarticle}
\usepackage{amsmath}
\usepackage{xspace}
\usepackage{graphicx}
\usepackage{lineno}
\usepackage{hyperref}

\usepackage{listings}
\usepackage{txfonts}  
\renewcommand{\vec}[1]{\mathbf{#1}}
\newcommand{\abs}[1]{\left|#1\right|}
\newcommand{\slashit}[1]{\ensuremath{{\not\mathrel{#1}}}}
\newcommand{\W}{\ensuremath{\mathrm{W}}\xspace}
\newcommand{\B}{\ensuremath{\mathrm{b}}\xspace}
\newcommand{\T}{\ensuremath{\mathrm{t}}\xspace}
\newcommand{\Q}{\ensuremath{\mathrm{q}}\xspace}

\title{Analytic solutions for neutrino momenta in decay of top quarks}
\author{Burton A.~Betchart\corref{cor1}}
\ead{bbetchar@pas.rochester.edu}
\cortext[cor1]{\url{bbetchar@pas.rochester.edu}}

\author{Regina Demina}
\ead{regina@pas.rochester.edu}

\author{Amnon Harel}
\ead{amnon.harel@cern.ch}

\address{Department of Physics and Astronomy, University of Rochester,
  Rochester, NY, United States of America}

\begin{document}

\begin{abstract}
  We employ a geometric approach to analytically solving equations of
  constraint on the decay of top quarks involving leptons.
  The neutrino momentum is found as a function of the 4-vectors of the
  associated bottom quark and charged lepton, the masses of the top
  quark and \W boson, and a single parameter, which constrains it to
  an ellipse.
  We show how the measured imbalance of momenta in the event reduces
  the solutions for neutrino momenta to a discrete set, in the cases
  of one or two top quarks decaying to leptons.
  The algorithms can be implemented concisely with common linear
  algebra routines.
\end{abstract}

\begin{keyword}
  top \sep neutrino \sep reconstruction \sep analytic
\end{keyword}

\maketitle

\section{Introduction}

Top quark reconstruction from channels containing one or more leptons
presents a challenge since the neutrinos are not directly observed.
The sum of neutrino momenta can be inferred from the total momentum
imbalance, but this quantity frequently has the worst resolution of
all constraints on top quark decays.
Reconstruction at hadron colliders faces further difficulties, since
the longitudinal momentum is unconstrained.
In a common approach to the single neutrino final state at hadron
colliders (e.g.~\cite{PhysRevD.85.051101}), constraining the invariant
mass of the neutrino and associated charged lepton to the \W boson
mass provides a quadratic equation for the unmeasured longitudinal
component of neutrino momentum, with zero solutions, or two solutions
which can be further resolved heuristically by consideration of
additional constraints.
We suggest an alternative approach to analytic top quark
reconstruction in which the invariant mass constraints from the top
quark and the \W boson are both exact, and in which the solution set
for each neutrino momentum is an ellipse.
For events with a single neutrino, the ellipse is analytically reduced
to a unique solution by application of the momentum imbalance
constraint, taking its uncertainty into account.
The approach extends naturally to the case of two neutrinos in the
final state, allowing an alternative method for calculating the
solution pairs previously described by
\cite{PhysRevD.45.1531,PhysRevD.72.095020,PhysRevD.73.054015}, and
suggesting a most likely pair in the case of no exact solution.

The solutions for the one- and two-neutrino cases are derived in
Section \ref{derivation}.
Their use in simulated Tevatron and LHC events, in the context of
iterative kinematic fit procedures, is discussed in Section
\ref{discussion}.
These results may also be useful for any event topology with similar
kinematic constraints, including decays involving new physics with
massive invisible particles, and hadronic decays of top quarks where
one of the quarks from the decay of the intermediate \W boson falls
outside experimental acceptance.

\section{Derivation}
\label{derivation}
The kinematics of top quark decay constrain the \W boson momentum
vector to an ellipsoidal surface of revolution about an axis
coincident with bottom quark momentum.
Simultaneously, the kinematics of \W boson decay constrain the \W
boson momentum vector to an ellipsoidal surface of revolution about an
axis coincident with the momentum of the resulting charged lepton.
The intersection of the two surfaces is an ellipse.
The neutrino momentum is consequently constrained to a translation of
the ellipse, for which a parametric expression in the laboratory
coordinate system is given.
The measured momentum imbalance further constrains solutions to a
discrete set for the cases of one or two top quark decays involving
neutrinos.

\subsection{Definitions}
A particle \Q is described by its mass $m_\Q$, energy $E_\Q$, and
momentum 3-vector $\vec{p}_\Q$, with the dispersion relation
\[m_\Q^2 = E_\Q^2 - \vec{p}_\Q^2.\]
The magnitude of the momentum is $p_\Q$.
Since there will be no need to denote positions, the Cartesian
coordinates of the momentum $\vec{p}_\Q$ in the laboratory coordinate
system are represented as $(x_\Q,y_\Q,z_\Q)$ in order to avoid double
subscripts.
In the laboratory coordinate system, the azimuthal and polar angles of
\Q are denoted $\phi_\Q$ and $\theta_\Q$, and the relativistic speed
and Lorentz factor are
\[
  \beta_\Q \equiv \frac{p_\Q}{E_\Q}, \qquad  \gamma_\Q^{-1} \equiv \frac{m_\Q}{E_\Q} = \sqrt{1-\beta_\Q^2}.
\]

For definiteness we consider the decay chain $\T\to\B\W\to\B\mu\nu$: a
top quark ($\T$) decays to a bottom quark (\B) and a \W boson (\W),
with subsequent decay of the \W boson to a muon ($\mu$) and a neutrino
($\nu$).
We assume established masses for all five particles, for example from
world average measurements\cite{PhysRevD.86.010001}.
Energy and momentum conservation for this system imply
\[
\begin{array}{c@{\quad=\quad}c@{\quad=\quad}c}
E_\T       & E_\B + E_\W           & E_\B + E_\mu + E_\nu, \\
\vec{p}_\T & \vec{p}_\B+\vec{p}_\W & \vec{p}_\B + \vec{p}_\mu+\vec{p}_\nu.
\end{array}
\]

\begin{figure}
  \centering
  \includegraphics[width=0.45\textwidth]{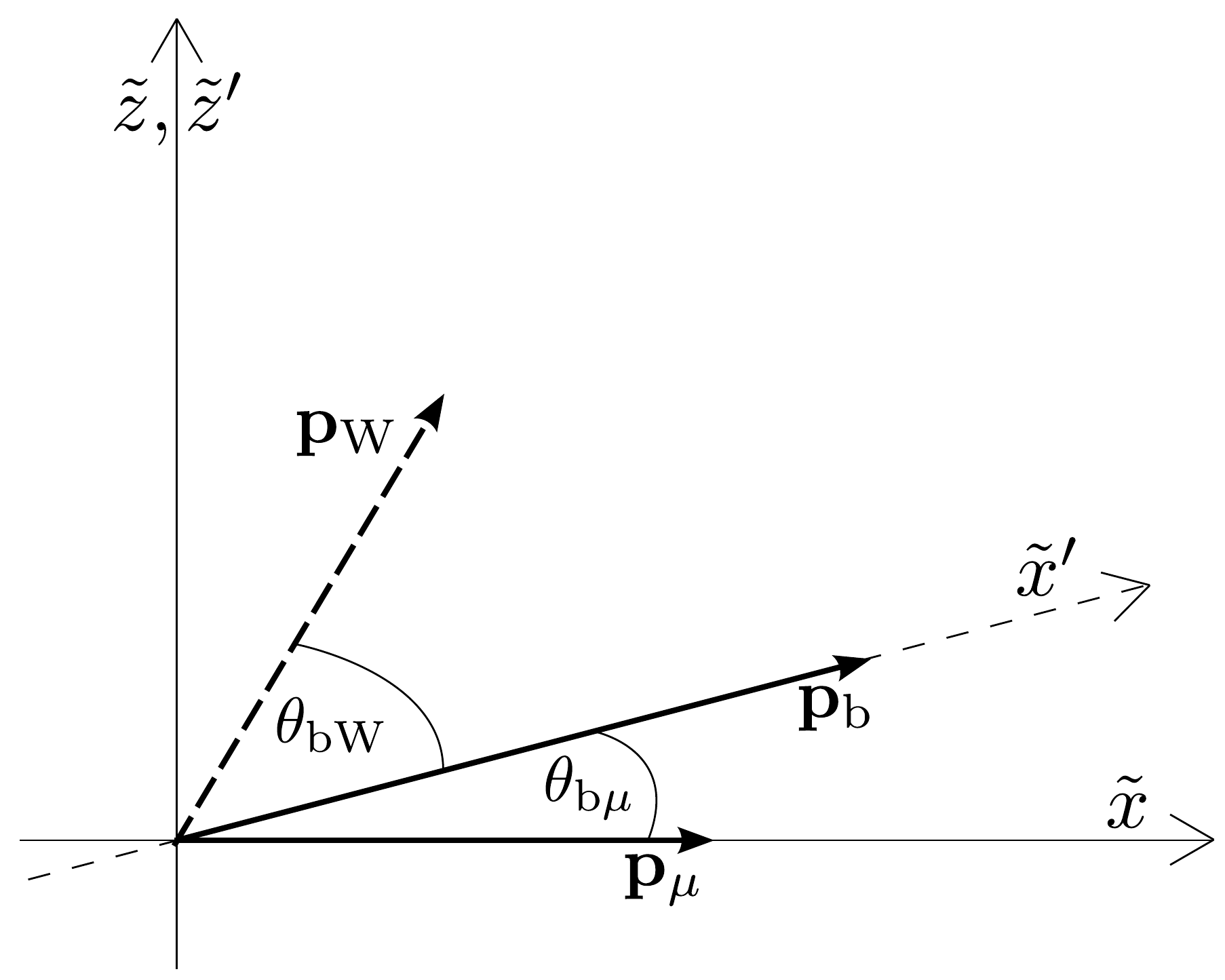}
  \caption{\label{fig_coordinates} %
    The momenta of observed particles $\mu$ and \B define the
    coordinate systems $F$ and $F'$.
    A possible momentum of the \W boson is drawn to show the angle
    $\theta_{\B\W}$.
  }
\end{figure}

Momentum coordinate systems $F\{\tilde{x},\tilde{y},\tilde{z}\}$
and $F'\{\tilde{x}',\tilde{y}',\tilde{z}'\}$ are defined in the
laboratory reference frame to share a common axis
$\tilde{z}=\tilde{z}'$.  Coordinate system $F'$ is rotated relative to
$F$ by the angle $\theta_{\B\mu}$ between $\vec{p}_\B$ and
$\vec{p}_\mu$, with $\vec{p}_\mu$ along the $\tilde{x}$-axis, and
$\vec{p}_\B$ along the $\tilde{x}'$-axis.
Polar and Cartesian coordinates of $\vec{p}_\W$ in $F'$ are related as
\begin{equation}
  \label{cosinesine}
  \tilde{x}'_\W = p_\W \mathcal{C}', \qquad
  \tilde{y}'^2_\W + \tilde{z}'^2_\W = p_\W^2\mathcal{S}'^2,
\end{equation}
where $\mathcal{S}'$($\mathcal{C}'$) is the (co)sine of the angle
$\theta_{\B\W}$ between $\vec{p}_\W$ and $\vec{p}_\B$.
Figure \ref{fig_coordinates} shows the coordinate systems.

\subsection{Two surfaces for \W momentum}

Energy and momentum are conserved in the decay $\T\to\B\W$, hence
\begin{eqnarray*}
  m_\T^2 &=& E_\T^2 - \vec{p}_\T^2\\
  &=& (E_\B+E_\W)^2 - (\vec{p}_\B+\vec{p}_\W)^2\\
  &=& m_\B^2 + m_\W^2 + 2E_\B E_\W - 2p_\B p_\W\mathcal{C'}.
\end{eqnarray*}
For compactness we define
\newcommand{\xO}{\tilde{x}_0}
\newcommand{\xOp}{\tilde{x}'_0}
\newcommand{\xxOp}{\tilde{x}'^2_0}
\begin{equation}
  \label{xnaughtprime_definition}
  \xOp \equiv - \frac{1}{2E_\B}\left(m_\T^2-m_\W^2-m_\B^2\right).
\end{equation}
It follows that
\begin{eqnarray*}
0 &=& \xOp + E_\W - \beta_\B p_\W\mathcal{C'} \\
0 &=& E_\W^2 - \left[\xOp - \beta_\B p_\W\mathcal{C'}\right]^2 \\
0 &=& m_\W^2 - \xxOp + 2\beta_\B p_\W\mathcal{C'}\xOp + p_\W^2\left(1 - \beta_\B^2\mathcal{C'}^2\right)   \\
0 &=& m_\W^2 - \xxOp + 2\beta_\B p_\W\mathcal{C'}\xOp + p_\W^2\left(\gamma_\B^{-2}\mathcal{C'}^2 + \mathcal{S'}^2\right).
\end{eqnarray*}
Incorporating relations (\ref{cosinesine}) for the $F'$ coordinates,
it is clear that $\vec{p}_\W$ is constrained to the surface
\begin{equation}
  \label{canon_ellipsoid}
  (\tilde{x}'/\gamma_\B)^2 + \tilde{y}'^2 + \tilde{z}'^2
  +2\beta_\B\xOp\tilde{x}' + \left(m_\W^2-\tilde{x}'^2_0\right) = 0,
\end{equation}
which is an ellipsoid of revolution about the $\tilde{x}'$-axis.

\begin{figure*}
  \centering
  \includegraphics[width=0.9\textwidth]{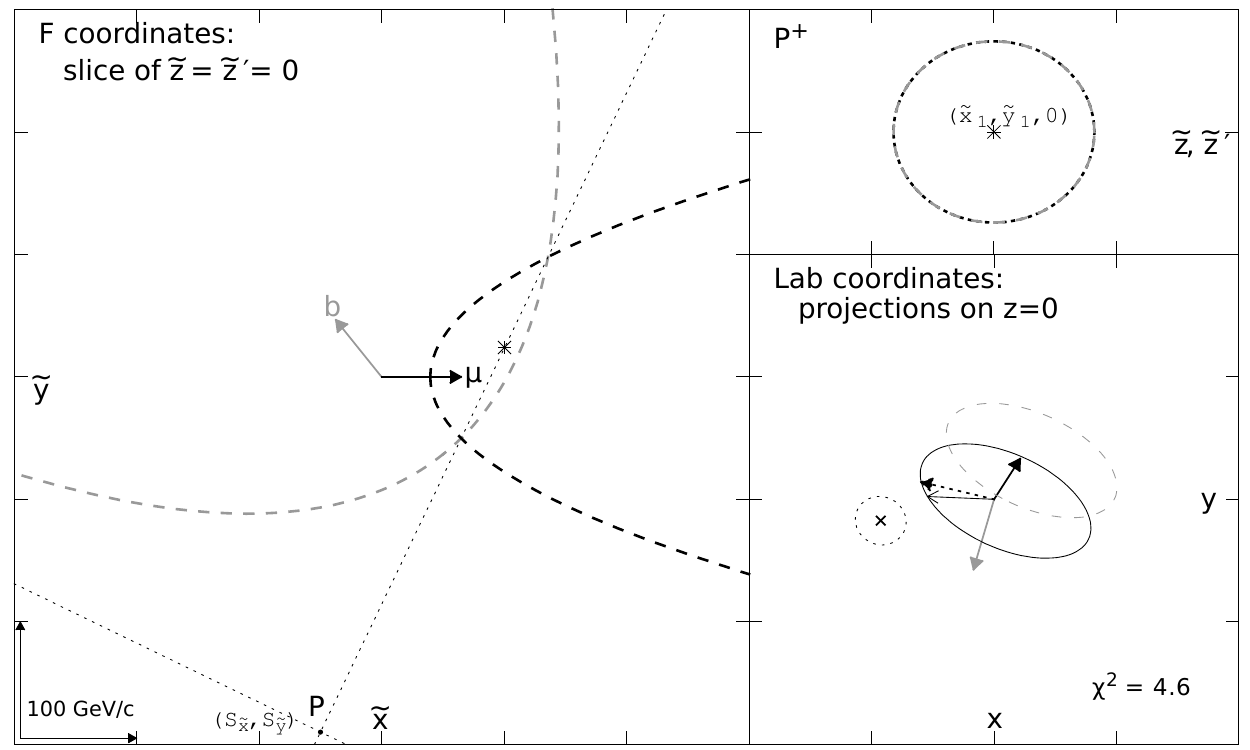}
  \includegraphics[width=0.9\textwidth]{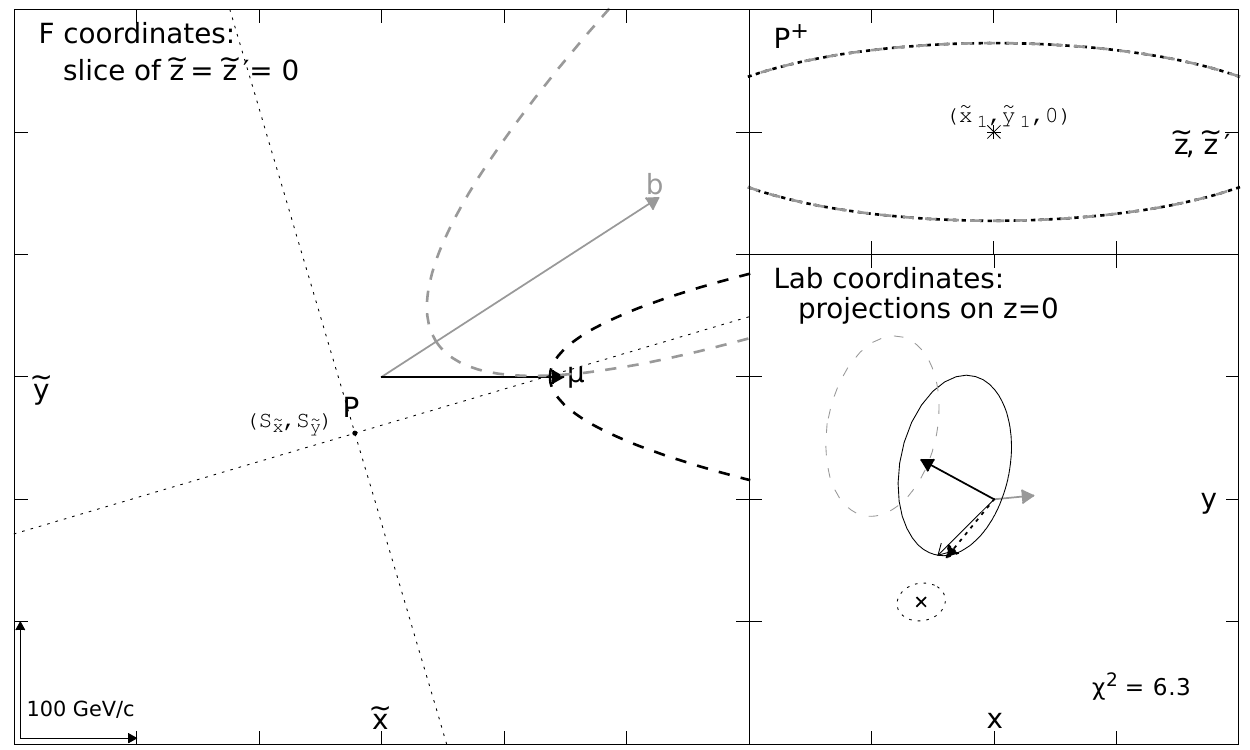}
  \caption{\label{fig_intersection}
    Reconstruction of neutrino momentum from the decay $\T\to\B\mu\nu$
    for two events, with $\theta_{\B\mu}$ large (top) and small
    (bottom).
    \textbf{Left} panel arrows indicate momenta of $\mu$ (black) and
    \B (gray).
    Corresponding constraints on the intermediate \W boson momentum
    are indicated (dashed) by their $\tilde{z}=0$ cross sections.
    The degenerate quadric surface $P$ (dotted) contains the
    intersection of the constraints.
    \textbf{Upper} panel ellipses show the intersection in the plane
    $P^+$.
    \textbf{Lower} panel filled arrows indicate laboratory transverse
    momenta of $\mu$ (black), \B (gray), and $\nu$ (dashed).
    The projection of the intersection ellipse from $P^+$ is traced
    in light gray dashes, while the corresponding ellipse of neutrino
    transverse momenta solutions is solid black.
    The measured imbalance in total transverse momentum and its
    uncertainty are shown by $\times$ and the encircling dots.
    The neutrino transverse momentum solution most compatible with the
    measurement is shown by the unfilled arrow.
  }
\end{figure*}

Particle \W subsequently decays to particles $\mu$ and $\nu$.  This
decay has the same kinematics as the decay $\T\to\B\W$, with the
substitutions $\B\to\mu$, $\W\to\nu$, and $\T\to\W$.
We define the \W decay analog of Equation
\ref{xnaughtprime_definition},
\[
  \xO \equiv - \frac{1}{2E_\mu}\left(m_\W^2-m_\mu^2-m_\nu^2\right).
\]
In analogy to Equation \ref{canon_ellipsoid} and using $F$
coordinates, $\vec{p}_\nu$ is constrained to the surface
\begin{equation}
  \label{nusurface}
  (\tilde{x}/\gamma_\mu)^2 + \tilde{y}^2 +\tilde{z}^2 + 2\beta_\mu\xO\tilde{x} + \left(m_\nu^2 - \xO^2\right) = 0,
\end{equation}
which is an ellipsoid of revolution about the $\tilde{x}$-axis.
A congruent surface of solutions for $\vec{p}_\W$ is translated from
the neutrino solutions (\ref{nusurface}) by $+p_\mu$ along the
$\tilde{x}$-axis, 
\newcommand{\Sx}{S_{\tilde{x}}}
\newcommand{\Sy}{S_{\tilde{y}}}
\begin{equation}
  \label{canon_paraboloid}
  (\tilde{x}/\gamma_\mu)^2 + \tilde{y}^2+\tilde{z}^2 +
  2\beta_\mu^2\Sx\tilde{x}
  +\left[m_\W^2 - \tilde{x}_0^2 - \epsilon^2\right] = 0,
\end{equation}
where for compactness and later use we have defined
\begin{equation}
  \label{Sx}
  \Sx = \left(\xO\beta_\mu - p_\mu\gamma_\mu^{-2}\right)/\beta_\mu^2,
\end{equation}
\[
  \epsilon^2 = \gamma_\mu^{-2}\left(m_\W^2-m_\nu^2\right).
\]
The solution set for $\vec{p}_\W$ is the intersection of two
simultaneous surfaces of constraint, (\ref{canon_ellipsoid}) and
(\ref{canon_paraboloid}), imposed by the \B measurement and the masses
($m_\T$, $m_\W$), and the $\mu$ measurement and the masses ($m_\W$,
$m_\nu$), respectively.
Figure \ref{fig_intersection} shows examples.
For relativistic particles $\mu$ and \B, the surfaces limit on
paraboloids, $\Sx$ limits on $\xO$, and $\epsilon^2$ limits on zero.

\subsection{Extended Matrix Representation}

The use of homogeneous coordinates $\vec{r}=(\ x\quad y\quad1\ )^T$ or
$\vec{s}=(\ x\quad y\quad z\quad 1\ )^T$ allows extended matrix
representation of various 2- or 3-dimensional geometric objects.
In particular, the $1\times3$ row matrix $L$ is the extended
representation of the line $L\vec{r}=0$ in two dimensions.
A $3\times3$ symmetric square matrix $M$ is the extended
representation of the conic section $\vec{r}^TM\vec{r}=0$ in two
dimensions.
A quadric surface in three dimensions, like a paraboloid or ellipsoid,
can be represented as a $4\times4$ symmetric square matrix $A$, with
$\vec{s}^TA\vec{s}=0$.
Extended representations are unique up to a multiplicative factor, and
allow transformations like rotations and translations to be expressed
by matrix multiplication\cite{chan:simple,homogeneous1,homogeneous2}.

The ellipsoid defined by particle \B (\ref{canon_ellipsoid}) is
represented for homogeneous $F'$ coordinates by the matrix
\[
  \tilde{A}'_{\B} = \left(\begin{array}{cccc}
    \gamma_\B^{-2}&0&0& \xOp\beta_\B \\
    0&1&0&0\\
    0&0&1&0\\
    \xOp\beta_\B & 0 & 0 & m_\W^2 - \xxOp
  \end{array}\right).
\]
The ellipsoid defined by particle $\mu$ (\ref{canon_paraboloid}) is
represented for homogeneous $F$ coordinates by the matrix
\[
  \tilde{A}_\mu = \left(\begin{array}{cccc}
    \gamma_\mu^{-2}&0&0&\Sx\beta_\mu^2\\
    0&1&0&0\\
    0&0&1&0\\
    \Sx\beta_\mu^2 &0&0& 
      m_\W^2-\xO^2 - \epsilon^2
  \end{array}\right).
\]
The transformation from homogeneous coordinates $F'$ to $F$ is given
by a rotation of $\theta_{\B\mu}$ around the $\tilde{z}$ axis,
\[
  K =
  \left(\begin{array}{cccc} 
    c & -s & 0 & 0 \\ 
    s & c & 0 & 0\\ 
    0&0&1&0 \\ 
    0&0 & 0 & 1
  \end{array}\right), 
  \begin{array}{r@{\ =\ }l} c & \cos\theta_{\B\mu},\\\quad s & \sin\theta_{\B\mu}.\end{array}
\]
The ellipsoid (\ref{canon_ellipsoid}) is represented in homogeneous
$F$ coordinates by the matrix $\tilde{A}_\B = K\tilde{A}'_\B K^T$,
\[
  \tilde{A}_\B = \left(\begin{array}{cccc}
    1 - (c\beta_\B)^2 & -cs\beta_\B^2 & 0 & c\xOp\beta_\B\\
    -cs\beta_\B^2 & 1 - (s\beta_\B)^2 & 0 & s\xOp\beta_\B\\
    0&0&1&0\\
    c\xOp\beta_\B & s\xOp\beta_\B & 0 & m_\W^2 - \xxOp
  \end{array}\right).
\]

\subsection{Intersections}

The pencil $P(\lambda)$ of quadrics $A$ and $A'$ is defined as
\[P(\lambda) 
= A' - \lambda A
= A\left(A^{-1}A' - \lambda I\right).\]
Levin provides two theorems relevant to intersections of quadrics (and
of conics)\cite{Levin:1976:PAD:360349.360355}.
First, if $A$ and $A'$ intersect, then all quadrics on their pencil
share the intersection.
Second, in the pencil of any two intersecting quadrics there exists a
ruled quadric, parametrized by a family of lines, which can be used
to parametrize the intersection curve.
Good candidates for the ruled quadric are singular quadrics, which
occur for $\lambda$ equal to any real eigenvalue of $A^{-1}A'$.

\subsubsection{Two ellipses}
\label{two_ellipses}
\begin{table*}
  \centering
   \begin{tabular}{rl|c}
     \hline
     \multicolumn{2}{c}{Case} & Factorization \\
     \hline
     horizontal \& vertical, & $G_{11}=0=G_{22}$    & $L_{+}=\left(\ G_{12}\quad0\quad G_{23}\ \right),\quad L_{-}=\left(\ 0\quad G_{12}\quad [G_{13}-G_{23}]\ \right)$\\
     parallel,               & $g_{33}=0\ne G_{22}$ & $L_{\pm} = \left(\begin{array}{ccc}G_{12} & G_{22} & \left[G_{23}\pm\sqrt{-g_{11}}\right]\end{array}\right)$\\
     intersecting,           & $G_{22}\ne0$         & $0 = \left( y - \frac{g_{23}}{g_{33}}\right)G_{22} +\left(x-\frac{g_{13}}{g_{33}}\right)\left(G_{12}\pm\sqrt{-g_{33}}\right)$\\
     \hline
  \end{tabular}
  \caption{\label{factorizations}
    Factorizations of the degenerate conic section
    $\protect\vec{r}^TG\protect\vec{r}=0$ into two lines
    $L_\pm\protect\vec{r}=0$, for various cases.
    The elements of $g$ are the cofactors of $G$.
    Numerical stability can be achieved by swapping indices
    $1\leftrightarrow2$ when necessary to enforce
    $\abs{G_{11}}<\abs{G_{22}}$, which also covers the cases
    $G_{22}=0$.}
\end{table*}

For later reference, we document the general solution for the
intersection of two coplanar ellipses, represented as $3\times3$
extended matrices $M$ and $M'$.
The degenerate conic $G=M-\lambda M'$, where $\lambda$ is a real
eigenvalue of $M^{-1}M'$, can be factored as a symmetrized outer
product of two lines
\[
  G=\left(L_+^T L_- + L_-^T L_+\right)/2.
\]
The factorizations for various conditions are listed in Table
\ref{factorizations}.
Points of intersection between a line $L\vec{r}=0$ and a conic
$\vec{r}^TM\vec{r}=0$ are eigenvectors of their cross product,
represented with Einstein summation convention as
\[
(L\times M)_{im} = \epsilon_{ijk}L_jM_{km}.
\]
Since the homogeneous coordinates are defined to have a value of 1 in
the third component, the eigenvectors must be scaled appropriately.
Note that the necessary eigenvalue computations are for $3\times3$
matrices, so the problem is equivalent to finding the roots of a cubic
function, for which analytic solutions are known\cite{cardano1968ars}.

\subsubsection{$\tilde{A}_\mu$ with $\tilde{A}_\B$}

By inspection, the quadric $P=\tilde{A}_\mu - \tilde{A}_\B$ is
singular, since it has no dependence on $\tilde{z}$.
Recall the definition of $\Sx$ (\ref{Sx}), and consider the
translation transformation $S$ given by
\[
  S = \left(\begin{array}{cccc} 
    1 & 0 & 0 & \Sx\\
    0 & 1 & 0 & \Sy\\
    0 & 0 & 1 & 0\\
    0 & 0 & 0 & 1
  \end{array}\right), \qquad
  \Sy = \frac{1}{s}\left(\xOp/\beta_\B - c\Sx \right).
\]
The translation of $P$ is
\[S^TPS = \beta_b^2\left(\begin{array}{cc}
  \begin{array}{cc}
    -\left(\beta_\mu/\beta_\B\right)^2 + c^2  & cs \\
    cs & s^2 \end{array} & \vec{0}_{2\times2}\\
  \vec{0}_{2\times2} & \vec{0}_{2\times2}
\end{array}\right),\]
which can be solved with the quadratic formula and translated back,
showing that $P=P^\pm$ is the pair of intersecting planes
\[
  \tilde{y} -\Sy = \omega(\tilde{x}-\Sx), \qquad \omega = \frac{1}{s}\left(\pm\frac{\beta_\mu}{\beta_\B}-c\right).
\]
Only $P^+$, the plane with positive slope, ever intersects
$\tilde{A}_\mu$.
In the $\tilde{z}=0$ plane the two points of intersection are
\[
  \left(\begin{array}{c} \tilde{x}_\pm \\ \tilde{y}_\pm \\ \tilde{z}_\pm \end{array}\right) = 
  \left(\begin{array}{c}
    \tilde{x}_1 \pm Z/\Omega\\      
    \tilde{y}_1 \pm \omega Z/\Omega\\
    0
  \end{array}\right),
\]
where
\begin{eqnarray*}
    \Omega^2    &=& \omega^2 + \gamma_\mu^{-2},\\
    \tilde{x}_1 &=& \Sx - (\Sx+\omega\Sy) / \Omega^2,\\
    \tilde{y}_1 &=& \Sy - (\Sx+\omega\Sy) \omega / \Omega^2,\\
    Z^2         &=& \tilde{x}_1^2\Omega^2 - \left(\Sy-\omega\Sx\right)^2 - \left( m_\W^2-\xO^2 - \epsilon^2 \right).
\end{eqnarray*}
For $Z^2=0$ the two points coincide, and $P^+$, $\tilde{A}_\B$ and
$\tilde{A}_\mu$ are tangent, while for $Z^2<0$ the constraints
$\tilde{A}_\mu$ and $\tilde{A}_\B$ are not consistent.
Since any cross section of an ellipsoid is an ellipse, $P^+$ cuts
$\tilde{A}_\B$ in an ellipse where it intersects $\tilde{A}_\mu$.
Since both $\tilde{A}_\B$ and $P^+$ are symmetric in $\pm\tilde{z}$,
the ellipse of intersection is symmetric in $\pm\tilde{z}$, so one
axis is in the plane $\tilde{z}=0$, between points $(\tilde{x}_\pm,
\tilde{y}_\pm, 0)$, and the other axis is perpendicular to
$\tilde{z}=0$, between points $(\tilde{x}_1, \tilde{y}_1, \pm Z)$.
This ellipse has $F$ coordinates parametrized by $t$,
\[
  \tilde{\vec{p}}_\W =
  \left(\begin{array}{c}
    \tilde{x}_1 + Z\cos(t)/\Omega \\
    \tilde{y}_1 + \omega Z\cos(t)/\Omega\\
    Z\sin(t)
  \end{array}\right).
\]
The corresponding ellipse of solutions for the neutrino momentum has
the $F$ coordinates $ \tilde{\vec{p}}_\nu = \tilde{H}\vec{t}$, where
\[
\tilde{H} = 
\left(\begin{array}{ccc}
  Z/\Omega &  0 & \tilde{x}_1-p_\mu  \\
  \omega Z/\Omega   &  0 & \tilde{y}_1        \\
  0   &  Z & 0
\end{array}\right),
\qquad \vec{t} = 
\left(\begin{array}{c}\cos t\\\sin t\\1\end{array}\right).
\]

\subsection{Laboratory coordinates}
The transformation of the laboratory coordinate system to the $F$
coordinate system can be accomplished by the following series of
rotations:
rotate the lab system around its $z$-axis by $\phi_\mu$, so that
$\vec{p}_\mu$ is in the $x'-z'$ plane, and $z'=z$, so all polar angles
are unchanged;
rotate the primed system around its $y'$-axis by
$\left(\theta_\mu-\frac{\pi}{2}\right)$, so that $\vec{p}_\mu$
coincides with $x''$;
rotate the double primed system around its $x''$-axis so that
$\vec{p}_\B$ is in the $\tilde{x}-\tilde{y}$ plane with
$\tilde{y}_\B>=0$.
The angle of the latter rotation, $\alpha$, is equal to the principal
value of the argument of ($y''_\B+z''_\B\sqrt{-1}$).
Noting that the rotation of the coordinate system is equivalent to the
opposite rotation of the vectors, the transformation from $F$
coordinates to laboratory coordinates is the rotation
\[
R =
R_z(\phi_\mu)
R_{y'}\left(\theta_\mu-\frac{\pi}{2}\right)
R_{x''}\left(\alpha\right).
\]
The set of neutrino momentum solutions is given in the laboratory
coordinates by the parametric form
\[
  \vec{p}_\nu = H\vec{t}, \qquad H = R\tilde{H}.
\]
For homogeneous coordinates in the transverse plane,
$\nu_\perp=(\ x_\nu\quad y_\nu\quad1\ )^T$ , the solutions are
\begin{equation}
  \label{transverse_parameterization}
  \nu_\perp = H_{\perp}\vec{t},\qquad
  H_{\perp} = \left(\begin{array}{ccc}H_{11}&H_{12}&H_{13}\\H_{21}&H_{22}&H_{23}\\0&0&1\end{array}\right)
\end{equation}
The extended representation of the solution ellipse in the transverse plane is
\begin{equation}
  \label{transverse_ellipse}
  N_{\perp} = H_{\perp}^{-T}UH_{\perp}^{-1},
\end{equation}
which follows trivially given that $\vec{t}$ parametrizes the
solution set of the unit circle
$U=\mathrm{diag}\left(\begin{array}{ccc}1&1&-1\end{array}\right)$.
Given a solution $\nu_\perp$, the full neutrino momentum is
\begin{equation}
  \label{full_from_transverse}
  \vec{p}_\nu = HH^{-1}_\perp\nu_\perp.
\end{equation}

\subsection{Momentum imbalance constraint}
\label{momentum_imbalance}
The solution sets of neutrinos from decaying top quarks can be further
constrained by the measured imbalance in momentum of the colliding
system $(\ \slashit{x}\quad\slashit{y}\quad\slashit{z}\ )$, which has
a diagonalizable $3\times3$ uncertainty matrix $\Sigma^2$.
We treat events with one or two top quarks decaying to leptons.

\subsubsection{Single neutrino in final state}
\label{single_neutrino}

The displacement between the measurement
$(\ \slashit{x}\quad\slashit{y}\quad\slashit{z}\ )$ and the solution
$\vec{p}_\nu=H\vec{t}$ is $\Lambda\vec{t}$, where
\[
  \label{displacement}
  \Lambda = V_0 - H, \qquad
  V_0=\left(\begin{array}{ccc}0&0&\slashit{x}\\0&0&\slashit{y}\\0&0&\slashit{z}\end{array}\right).
\]
The weighted square of the displacement is
\[
  \chi^2 = \vec{t}^TX\vec{t},
  \qquad X = \Lambda^T\Sigma^{-2}\Lambda.
\]
Note that systems with unconstrained longitudinal momentum can be
accommodated by setting the upper $2\times2$ submatrix of
$\Sigma^{-2}$ to the inverse of the $2\times2$ uncertainty matrix of
the transverse momentum imbalance, and the rest of the entries and
$\slashit{z}$ to zero.

We seek the neutrino solution with the minimum value of $\chi^2$.
Differentiation of $\vec{t}$ with respect to $t$ can be expressed as
matrix multiplication,
\[
\frac{\partial \vec{t}}{\partial t} = D\vec{t},
\qquad D = \left(\begin{array}{ccc}0&-1&0\\1&0&0\\0&0&0\end{array}\right).
\]
The extrema of $\chi^2$ occur at values of $t$ such that
\[
  \frac{\partial\chi^2}{\partial t} = \vec{t}^TM\vec{t} = 0, \qquad M = (XD)^T+XD.
\]
Since $M$ is symmetric, $\vec{t}$ is a point on the conic described by
$M$.
Since $\vec{t}$ is also a point on the unit circle, solutions must be
points on the intersection of $M$ and
$U=\mathrm{diag}\left(\ 1\quad1\quad-1\ \right)$, which can be found
using the method documented in Section \ref{two_ellipses}.
We expect at least one minimum of $\chi^2$.
Since $\chi^2$ is cyclic, no more than 2 of at most 4 intersections
can be minima, one of which is the global minimum we seek.

\subsubsection{Two neutrinos in final state}

\begin{figure}
  \centering
  \includegraphics[width=0.328\textwidth]{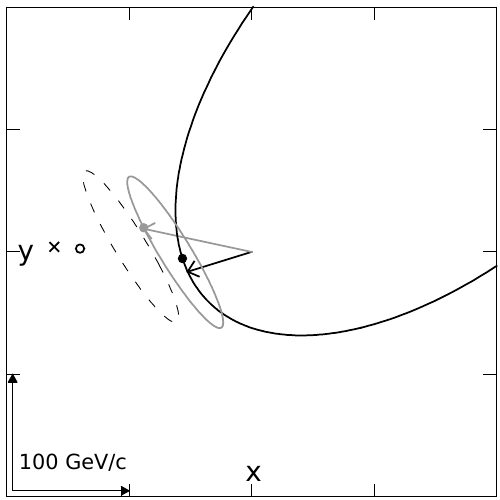}
  \includegraphics[width=0.328\textwidth]{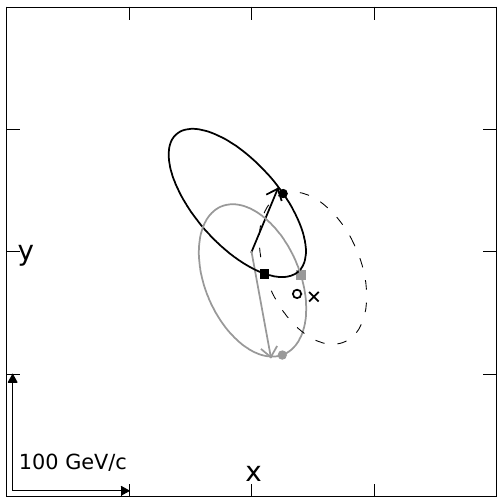}
  \includegraphics[width=0.328\textwidth]{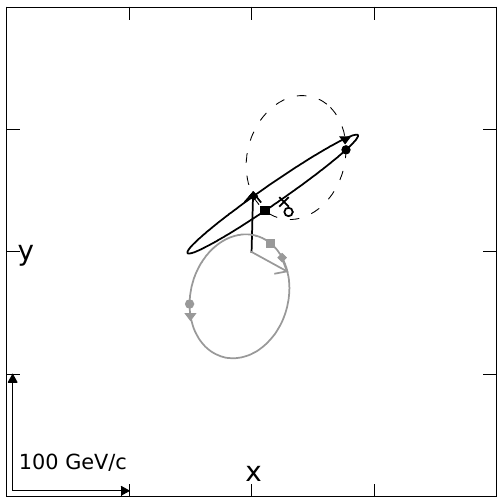}
  \caption{\label{fig:dilepton}
    Constraints on neutrino momenta from the decay of two top quarks,
    in three events.
    Arrows indicate the neutrino (black) and antineutrino (gray)
    laboratory transverse momenta, which are constrained to respective
    ellipses (solid) by the associated products of top quark decays.
    The dashed ellipse is an additional constraint on the neutrino
    momentum from the antineutrino constraints, under the hypothesis
    that the measured imbalance in transverse momenta ($\times$) is
    equal to the sum of neutrino transverse momenta ($\circ$).
    Exact solutions occur at intersections of the black and dashed
    ellipses, and can number zero (first), two (center), or four
    (last).
    With no intersection (first), the single best solution pair is for
    points of closest approach on the black and dashed ellipses.
    Solution pairs are marked in matching shaped points on the
    respective ellipses.
}
\end{figure}

Suppose that the longitudinal momentum of the system is unconstrained.
Given two top quarks decaying to leptons, $\T\to\B\nu\mu^+$ and
$\bar{t}\to\bar{\B}\bar{\nu}\mu^-$, the respective elliptical solution
sets for neutrino transverse momenta are given by Equation
(\ref{transverse_ellipse}),
\[\nu_\perp^TN_\perp\nu_\perp = 0, \qquad \bar{\nu}_\perp^T\bar{N}_\perp\bar{\nu}_\perp = 0.\]
Since the measured components $(\ \slashit{x}\quad\slashit{y}\ )$ of
transverse momentum imbalance are ideally just the sum of $\nu_\perp$
and $\bar{\nu}_\perp$ components, they are related by
\begin{equation}
  \label{sum_constraint}
  \bar{\nu}_{\perp} =  \Gamma\nu_\perp, \qquad \Gamma = \left(\begin{array}{ccc}-1&0&\slashit{x}\\0&-1&\slashit{y}\\0&0&1\end{array}\right).
\end{equation}
Note that $\Gamma=\Gamma^{-1}$.
We can rewrite the $\bar{\nu}_\perp$ ellipse in terms of $\nu_\perp$,
\begin{equation}
  \label{nubar_ellipse}
  \nu_\perp^T\bar{N}'_\perp\nu_\perp = 0, \qquad \bar{N}'_\perp = \Gamma^T\bar{N}_\perp \Gamma.
\end{equation}
Solutions for $\nu_\perp$ are on both ellipses $N_{\perp}$ and
$\bar{N}'_{\perp}$ simultaneously, and can be found using the method
documented in Section \ref{two_ellipses}.
There can be zero, two, or four intersections, discounting cases of
tangency.
Each intersection implies a solution pair ($\vec{p}_\nu$,
$\vec{p}_{\bar{\nu}}$), which can be found using Equations
(\ref{full_from_transverse}) and (\ref{sum_constraint}).
Examples are shown in Figure \ref{fig:dilepton}.
These solution pairs are identical to those described in Reference
\cite{PhysRevD.73.054015}, which contains a detailed discussion of
solution efficiency and multiplicity under various collider, detector,
and combinatorial scenarios.

If $N_\perp$ and $\bar{N}'_\perp$ do not intersect, Equation
(\ref{nubar_ellipse}) is never satisfied, so the points of closest
approach constitute the single solution pair most likely to result in
the observed transverse momentum imbalance.
In the case that the ellipse $\bar{N}'_\perp$ has low eccentricity,
the closest approach of $\nu_\perp$ is well-approximated by the
extremum closest to zero of the function
\[
  f = \nu_\perp^T \bar{N}'_\perp \nu_\perp = \vec{t}^T X'\vec{t},\qquad X'=H_\perp^T \bar{N}'_\perp H_\perp,
\]
where we have used the parametrization
(\ref{transverse_parameterization}).
Extrema of $f$ occur for
\[
  \frac{\partial f}{\partial t} = \vec{t}^TM'\vec{t} = 0,\qquad M' = (X'D)^T + X'D,
\]
which as in Section \ref{single_neutrino} is solved for $\vec{t}$ on
intersections of the unit circle $U$ with the conic described by $M'$.
Once $\nu_\perp$ is known, the closest approach on $\bar{N}'_\perp$ is at
one of its intersections with the line perpendicular to $N_\perp$ at
$\nu_\perp$, given by $\nu_\perp\times(D^2N_\perp\nu_\perp)$.
Alternatively, least squares minimization of the distance between
points on the ellipses is straightforward to implement using
parametrization (\ref{transverse_parameterization}), which also
facilitates incorporation of the uncertainty.

In the case that the longitudinal momentum is constrained, an
analogous strategy can be followed without projecting the elliptical
solution sets onto the transverse plane.
The analogous ellipses $N$ and $\bar{N}'$ are not generally coplanar
or intersecting, implying a single solution pair of weighted closest
approach which can be found iteratively.

\section{Discussion}
\label{discussion}
In the overconstrained problems of Section \ref{momentum_imbalance},
for which the solutions cannot be exact, there are additional sources
of uncertainty beyond the total momentum imbalance.
In particular, the resolution of jet momenta is significant in many
experiments.
Many successful iterative methods have been developed to find the smallest
corrections to the measured momenta such that the invariant mass
constraints are satisfied, by minimization of squared residuals
\cite{Snyder:1995hg, KinFitter}, or alternatively by maximization of
the likelihood\cite{KLFitter, Loddenkotter:2012eca}.
Iterative kinematic fit methods have been widely used in top quark
analyses (e.g.~\cite{PhysRevD.75.092001,JHEP.12.105}).
We suggest that the overconstrained analytic solutions of Section
\ref{momentum_imbalance} be used in conjunction with iterative methods.
Incorporating the solutions into a broader iterative algorithm allows
parameters associated with neutrino momenta to be removed from the
global fit, while preserving the associated residuals or contributions
to the likelihood.
We found that utilization of the overconstrained analytic neutrino
solutions improves the convergence speed and reliability of
constrained least squares algorithms for the decay of
$\mathrm{t\bar{t}}$ to one charged lepton and jets at the LHC and at
the Tevatron.
The amount of improvement is dependent on implementation details and
resolution characteristics of each detector, the discussion of which
is beyond the scope of this paper.

\section{Summary}

The momentum of a top quark is constrained to an ellipsoidal surface
defined by its invariant mass, the \W boson invariant mass, and the
observable kinematic properties of the bottom quark produced in its
decay.
Its momentum is further constrained to a slice of that surface by the
\W boson invariant mass and the observed kinematic properties of
either of the products from the decay of the intermediate \W boson.
The other decay product of the \W is likewise constrained to an
ellipse, for which we have given a parametrization.
In the event that a single neutrino is produced from top quark decay,
we have calculated the momentum on its elliptical solution set which
is most likely to have produced the observed momentum imbalance.
In the event that two neutrinos are produced from distinct top quark
decays, we have shown a method to calculate the discrete solution set
of momentum pairs which exactly produce the observed transverse
momentum imbalance, or the solution pair most likely to have produced
the momentum imbalance if exact solutions do not exist.
Due to the geometric nature of the constraints, their description is
facilitated by the use of homogeneous coordinates and extended matrix
representations.
Algorithms for computing these solutions can be concisely implemented
using common linear algebra routines.
We implemented such algorithms in Python and also in C++, and tested
for LHC and Tevatron use cases.
A reference implementation is included as an appendix.

\section*{Acknowledgments}

The authors acknowledge support from the Department of Energy under
the grant DE-SC0008475.

\bibliography{references.bib}{}
\bibliographystyle{model1-num-names}

\appendix
\section{Python Reference Implementation}
\begin{lstlisting}[language=Python]
import numpy as np
import ROOT as r
import math
from scipy.optimize import leastsq


mT = 172.5   # GeV : top quark mass
mW = 80.385  # GeV : W boson mass
mN = 0       # GeV : neutrino mass


def UnitCircle():
    '''Unit circle in extended representation'''
    return np.diag([1, 1, -1])


def cofactor(A, (i, j)):
    '''Cofactor[i,j] of 3x3 matrix A'''
    a = A[not i:2 if i==2 else None:2 if i==1 else 1,
          not j:2 if j==2 else None:2 if j==1 else 1]
    return (-1)**(i+j) * (a[0,0]*a[1,1] - a[1,0]*a[0,1])


def R(axis, angle):
    '''Rotation matrix about x(0),y(1), or z(2) axis'''
    c, s = math.cos(angle), math.sin(angle)
    R = c * np.eye(3)
    for i in [-1, 0, 1]:
        R[(axis-i)%3, (axis+i)%3] = i*s + (1 - i*i)
    return R


def Derivative():
    '''Matrix to differentiate [cos(t),sin(t),1]'''
    return R(2, math.pi / 2).dot(np.diag([1, 1, 0]))


def multisqrt(y):
    '''Valid real solutions to y=x*x'''
    return ([] if y < 0 else
            [0] if y == 0 else
            (lambda r: [-r, r])(math.sqrt(y)))


def factor_degenerate(G, zero=0):
    '''Linear factors of degenerate quadratic polynomial'''
    if G[0,0] == 0 == G[1,1]:
        return [[G[0,1], 0, G[1,2]],
                [0, G[0,1], G[0,2] - G[1,2]]]

    swapXY = abs(G[0,0]) > abs(G[1,1])
    Q = G[(1,0,2),][:,(1,0,2)] if swapXY else G
    Q /= Q[1,1]
    q22 = cofactor(Q, (2,2))

    if -q22 <= zero:
        lines = [[Q[0,1], Q[1,1], Q[1,2]+s]
                 for s in multisqrt(-cofactor(Q, (0,0)))]
    else:
        x0, y0 = [cofactor(Q,(i,2)) / q22 for i in [0, 1]]
        lines = [[m, Q[1,1], -Q[1,1]*y0 - m*x0]
                 for m in [Q[0,1] + s
                           for s in multisqrt(-q22)]]

    return [[L[swapXY],L[not swapXY],L[2]] for L in lines]


def intersections_ellipse_line(ellipse, line, zero=1e-12):
    '''Points of intersection between ellipse and line'''
    _,V = np.linalg.eig(np.cross(line,ellipse).T)
    sols = sorted([(v.real / v[2].real,
                    np.dot(line,v.real)**2 +
                    np.dot(v.real,ellipse).dot(v.real)**2)
                   for v in V.T],
                  key=lambda (s, k): k)[:2]
    return [s for s, k in sols if k < zero]


def intersections_ellipses(A, B, returnLines=False):
    '''Points of intersection between two ellipses'''
    LA = np.linalg
    if abs(LA.det(B)) > abs(LA.det(A)): A,B = B,A
    e = next(e.real for e in LA.eigvals(LA.inv(A).dot(B))
             if not e.imag)
    lines = factor_degenerate(B - e*A)
    points = sum([intersections_ellipse_line(A,L)
                  for L in lines],[])
    return (points,lines) if returnLines else points


class nuSolutionSet(object):
    '''Definitions for nu analytic solution, t->b,mu,nu'''

    def __init__(self, b, mu,  # Lorentz Vectors
                 mW2=mW**2, mT2=mT**2, mN2=mN**2):
        c = r.Math.VectorUtil.CosTheta(b,mu)
        s = math.sqrt(1-c**2)

        x0p = - (mT2 - mW2 - b.M2()) / (2*b.E())
        x0 = - (mW2 - mu.M2() - mN2) / (2*mu.E())

        Bb, Bm = b.Beta(), mu.Beta()

        Sx = (x0 * Bm - mu.P()*(1-Bm**2)) / Bm**2
        Sy = (x0p / Bb - c * Sx) / s

        w = (Bm / Bb - c) / s
        w_ = (-Bm / Bb - c) / s

        Om2 = w**2 + 1 - Bm**2
        eps2 = (mW2 - mN2) * (1 - Bm**2)
        x1 = Sx - (Sx+w*Sy) / Om2
        y1 = Sy - (Sx+w*Sy) * w / Om2
        Z2 = x1**2 * Om2 - (Sy-w*Sx)**2 - (mW2-x0**2-eps2)
        Z = math.sqrt(max(0, Z2))

        for item in ['b','mu','c','s','x0','x0p',
                     'Sx','Sy','w','w_','x1','y1',
                     'Z','Om2','eps2','mW2']:
            setattr(self, item, eval(item))

    @property
    def K(self):
        '''Extended rotation from F' to F coord.'''
        return np.array([[self.c, -self.s, 0, 0],
                         [self.s,  self.c, 0, 0],
                         [     0,       0, 1, 0],
                         [     0,       0, 0, 1]])

    @property
    def A_mu(self):
        '''F coord. constraint on W momentum: ellipsoid'''
        B2 = self.mu.Beta()**2
        SxB2 = self.Sx * B2
        F = self.mW2 - self.x0**2 - self.eps2
        return np.array([[1-B2, 0, 0, SxB2],
                         [   0, 1, 0,    0],
                         [   0, 0, 1,    0],
                         [SxB2, 0, 0,    F]])

    @property
    def A_b(self):
        '''F coord. constraint on W momentum: ellipsoid'''
        K, B = self.K, self.b.Beta()
        mW2, x0p = self.mW2, self.x0p
        A_b_ = np.array([[1-B*B,  0,  0,      B*x0p],
                         [    0,  1,  0,          0],
                         [    0,  0,  1,          0],
                         [B*x0p,  0,  0, mW2-x0p**2]])
        return K.dot(A_b_).dot(K.T)

    @property
    def R_T(self):
        '''Rotation from F coord. to laboratory coord.'''
        b_xyz = self.b.x(), self.b.y(), self.b.z()
        R_z = R(2, -self.mu.phi())
        R_y = R(1, 0.5*math.pi - self.mu.theta())
        R_x = next(R(0,-math.atan2(z,y))
                   for x,y,z in (R_y.dot(R_z.dot(b_xyz)),))
        return R_z.T.dot(R_y.T.dot(R_x.T))

    @property
    def H_tilde(self):
        '''Transformation of t=[c,s,1] to p_nu: F coord.'''
        x1, y1, p = self.x1, self.y1, self.mu.P()
        Z, w, Om = self.Z, self.w, math.sqrt(self.Om2)
        return np.array([[  Z/Om, 0, x1-p],
                         [w*Z/Om, 0,   y1],
                         [     0, Z,    0]])

    @property
    def H(self):
        '''Transformation of t=[c,s,1] to p_nu: lab coord.'''
        return self.R_T.dot(self.H_tilde)

    @property
    def H_perp(self):
        '''Transformation of t=[c,s,1] to pT_nu: lab coord.'''
        return np.vstack([self.H[:2], [0, 0, 1]])

    @property
    def N(self):
        '''Solution ellipse of pT_nu: lab coord.'''
        HpInv = np.linalg.inv(self.H_perp)
        return HpInv.T.dot(UnitCircle()).dot(HpInv)


class singleNeutrinoSolution(object):
    '''Most likely neutrino momentum for tt-->lepton+jets'''
    def __init__(self, b, mu,   # Lorentz Vectors
                 (metX, metY),  # Momentum imbalance
                 sigma2,        # Mo. imbalance unc. matrix
                 mW2=mW**2, mT2=mT**2):
        self.solutionSet = nuSolutionSet(b, mu, mW2, mT2)
        S2 = np.vstack([np.vstack([np.linalg.inv(sigma2),
                                   [0, 0]]).T, [0, 0, 0]])
        V0 = np.outer([metX, metY, 0], [0, 0, 1])
        deltaNu = V0 - self.solutionSet.H

        self.X = np.dot(deltaNu.T, S2).dot(deltaNu)
        M = next(XD + XD.T
                 for XD in (self.X.dot(Derivative()),))

        solutions = intersections_ellipses(M, UnitCircle())
        self.solutions = sorted(solutions, key=self.calcX2)

    def calcX2(self, t):
        return np.dot(t, self.X).dot(t)

    @property
    def chi2(self):
        return self.calcX2(self.solutions[0])

    @property
    def nu(self):
        '''Solution for neutrino momentum'''
        return self.solutionSet.H.dot(self.solutions[0])


class doubleNeutrinoSolutions(object):
    '''Solution pairs of neutrino momenta, tt -> leptons'''
    def __init__(self, (b, b_), (mu, mu_),  # 4-vectors
                 (metX, metY),              # ETmiss
                 mW2=mW**2, mT2=mT**2):
        self.solutionSets = [nuSolutionSet(B, M, mW2, mT2)
                             for B,M in zip((b,b_),(mu,mu_))]

        V0 = np.outer([metX, metY, 0], [0, 0, 1])
        self.S = V0 - UnitCircle()

        N, N_ = [ss.N for ss in self.solutionSets]
        n_ = self.S.T.dot(N_).dot(self.S)

        v = intersections_ellipses(N, n_)
        v_ = [self.S.dot(sol) for sol in v]

        if not v and leastsq:
            es = [ss.H_perp for ss in self.solutionSets]
            met = np.array([metX, metY, 1])

            def nus(ts):
                return tuple(e.dot([math.cos(t), math.sin(t), 1])
                             for e, t in zip(es, ts))

            def residuals(params):
                return sum(nus(params), -met)[:2]

            ts,_ = leastsq(residuals, [0, 0],
                           ftol=5e-5, epsfcn=0.01)
            v, v_ = [[i] for i in nus(ts)]

        for k, v in {'perp': v, 'perp_': v_, 'n_': n_}.items():
            setattr(self, k, v)

    @property
    def nunu_s(self):
        '''Solution pairs for neutrino momenta'''
        K, K_ = [ss.H.dot(np.linalg.inv(ss.H_perp))
                 for ss in self.solutionSets]
        return [(K.dot(s), K_.dot(s_))
                for s, s_ in zip(self.perp, self.perp_)]
\end{lstlisting}

\end{document}